\RequirePackage{fix-cm}
\documentclass[smallextended]{svjour3}       
\smartqed  
\usepackage{graphicx}
\usepackage{epstopdf}

\begin{document}

\title{Mathematical models describing the effects\\
of different tax evasion behaviors
}


\author{M.L.  Bertotti \and G. Modanese}

\authorrunning{Short form of author list} 

\institute{M.L. Bertotti \at
                Faculty of Science and Technology, Free University of Bozen-Bolzano, Piazza Universit\`a 5, 39100 Bolzano, ITALY\\
              \email{MariaLetizia.Bertotti@unibz.it}          
               \and
               G. Modanese \at
               Faculty of Science and Technology, Free University of Bozen-Bolzano, Piazza Universit\`a 5, 39100 Bolzano, ITALY\\
               \email{Giovanni.Modanese@unibz.it}         
           }

\date{Received: date / Accepted: date}

\maketitle

\begin{abstract}
Microscopic models describing a whole of economic interactions in a closed society are  considered. 
The presence of a tax system
combined with a redistribution process is taken into account,
as well as the occurrence of tax evasion.
In particular, the existence is postulated, in relation to the level of evasion,
of different individual taxpayer behaviors. 
The effects of the mentioned different behaviors
on shape and features of the emerging income distribution profile
are investigated
qualitatively and quantitatively.
Numerical solutions show that the Gini inequality index of the total population increases when the evasion level is higher, but does not depend significantly on the evasion spread. For fixed spread, the relative difference between the average incomes of the worst evaders and honest taxpayers increases approximately as a quadratic function of the evasion level.
\end{abstract}

\section{Introduction}
\label{intro}

The phenomenon of tax evasion represents a serious problem for several countries. 
Having as a main consequence
a reduction of the tax revenue, it negatively affects 
a correct functioning of the public sector: it hurts
the supply of education, health care and services in general.
In turn,
this contributes, together with other factors, to an increase of economic inequality.

In this paper, we try and look at some aspects of
the problem through a mathematical-modelling approach.
Specifically, we discuss a kinetic-type model for economic exchanges in a closed society in the presence of
taxation and redistribution, within which occurrence of tax evasion to various extents is assumed:
we admit the possibility that different citizens pay different percentages of the taxes they should pay. 

We emphasize that the illegal practice under consideration
involves in fact a large number of interacting agents.
These include clearly the evaders themselves, but in addition also all other citizens who, 
by the situation, are deprived of the access to the benefits deriving from the revenue redistribution. 
In view of this and of the various levels at which it can take place, tax evasion can be thought of as 
an example of complex system. 
With this denomination, systems composed by a high number of heterogeneous units
are meant, whose collective and macroscopic behavior 
is not derivable from the simple summation of the single units 
properties, but inherently depends on
their nonlinear interplay.
An interaction-based approach 
seems to be quite a natural one
in the study of economic questions,
but it was only during the last decades,
especially thanks to the new opportunities offered by the increased 
computer power, that it started to be pursued. 
Arguments in favour and related work can be found e.g. in
\cite{RefAoki,RefArthur,RefGalKir,RefKir,RefKirman,RefLanGalSti,RefTesJud}.
The technical tools 
more frequently employed in connection with this approach
are agent-based computational algorithms
and simulations, possibly combined with complex networks theory.
Also, starting in the mid-1990s an interdisciplinary research field 
called econophysics has been growing, 
which explores the dynamical behaviour of economic and financial markets by 
means of methods taken from
statistical mechanics and gas kinetic theory,
see e.g. \cite{RefChaCha,RefChaYarCha,RefHeiPat,RefYak}.

Adopting here a mathematics-supported complex system perspective, 
we aim at deriving and explaining
the emergence of a population's aggregate feature like the income distribution,
as a result of the whole of economic exchanges and interactions which take place between the individuals.
More specifically, our focus here is on the effects that the heterogeneity of
taxpayer behaviors has on the income distribution profiles
of the categories of individuals evading to different degrees. To this end we build on our previous work,
incorporating suitable additional facets
into the models discussed in \cite{RefB,RefBM1,RefBM2,RefBM3,RefBM4}.
We also emphasize that our goal here is mainly
a methodological one: 
more than perfectly representing real world features,
we aim at constructing a tool endowed with explorative ability.
Our model in fact has this character. Indeed, the
possibility of finding numerical solutions which evolve from differently chosen ``initial conditions"
and in the presence of differently tuned parameters
amounts to the ability to
forecast 
the emergence of different scenarios.
In turn, this 
can give insights as to which policies 
could be adopted
to favour or prevent desired and undesired trends.

The tax evasion process is the object of a large amount of literature.
Works among those which involve agent based modeling and simulations
include e.g.
\cite{RefBlo,RefCro,RefHokSei,RefZakWesSta}, to name but a few. 
Specific aspects therein investigated concern
the occurrence of behavioral changes of agent types with diverse moral attitude, due to imitation or also to some audit procedure.
In particular, in \cite{RefHokSei} and \cite{RefZakWesSta} a variant of the Ising model originally developed within the theory of magnetism is considered.
In that context, each spin represents a citizen, 
which can be either in the tax compliant state $+1$ or in the tax evader state $-1$. 
Citizens undergo transitions from $+1$ to $-1$ caused by imitation of their nearest neighbours, and
from $-1$ to $+1$ induced by tax audits. Indeed, the consequence of an audit on an evader is assumed to 
be the fact that she/he will remain honest for a certain number of steps.
This approach is helpful for the analysis of evasion phenomena in relation to local interaction
and external controls, but not for studying the effect of evasion on the income distribution as we do here.

The paper is organised as follows. In Section \ref{model} we describe the model and some of its analytical properties. 
In Section \ref{numsim} we report and discuss results from a set of numerical solutions. 
Section \ref{conclud} contains our conclusions.

\section{The model}
\label{model}

This section is devoted to a short description of the model proposed. 
More details on the primary mechanism underlying it
can be found in our papers \cite{RefB,RefBM1,RefBM2,RefBM3,RefBM4}. 
We point out however
that in \cite{RefB,RefBM1,RefBM2} the phenomenon of tax evasion was not taken into account,
whereas in \cite{RefBM3,RefBM4} all individuals were assumed to engage to the same degree
in a kind of evasion different from that dealt with here.\footnote{\ In \cite{RefBM3,RefBM4} a kind of evasion was considered,
which provides advantage to both the participants in a transaction, as it 
sometimes happens in relation to value added taxes. Here, we consider evasion 
of individuals who under-declare their income.}
The main novelty here is given by the assumption of the existence of different degrees of evasion.

The totality of individuals (supposed to remain constant in time, in fact
during the time period under consideration) is divided 
into a number $n$ of {\it {classes}}, each one characterized by its average income,
the average incomes being the positive numbers $r_1< r_2 < ... <\ r_n$,
and, in turn, every income class is divided into a number $m$
of {\it {sectors}} characterized by possible evasion behaviors. 

We denote by $x_j^{\alpha}$ the fraction of individuals belonging 
to the $j$-th income class and to the $\alpha$-th evasion behavior sector.
We will say for brevity that $x_j^{\alpha}$ is the fraction of individuals of type $(j,\alpha)$,
also called $(j,\alpha)$-individuals, the number of different groups being $n \times m$.

We suppose here that the evasion behavior of each individual remains constant in time.
In contrast, individuals may move through different income classes.
The model is formulated by a system of $n \times m$ ordinary differential equations 
describing the variation in time of $x_j^{\alpha}$ for $j = 1,...,n$ and $\alpha = 1,..,m$.
Such a variation is the result of direct economic interactions, in which pairs of individuals exchange some money. 
And is affected as well by the payment of taxes (in some cases, the due taxes and in other cases,
partial quotes of them) and by the revenue redistribution, 
represented in the real world by healthcare, education and services in general. 
More specifically, the dynamic process is as follows:
a whole of interactions between pairs of individuals occur simultaneously:
for any $h$ and $k$ in $\{1,...,n\}$, for any $\beta$ and $\gamma$ in $\{1,...,m\}$
individuals belonging 
to the $h$-th income class and the $\beta$-th evasion sector meet individuals 
of the $k$-th income class and the $\gamma$-th evasion sector
and some money exchange between such pairs takes place.
How many of these economic exchanges do we have? If at the considered time the fraction of $(h,\beta)$-individuals is $x_{h}^{\beta}$  
and
the fraction of $(k,\gamma)$-individuals is $x_{k}^{\gamma}$, the number of encounters of these two categories of individuals is the product $x_{h}^{\beta} x_{k}^{\gamma}$.
And any single encounter contributes, albeit to a very small extent, to a change of the fraction of individuals in some income classes.
The 
differential
equations 
contain several parameters, 
which express for example transition probabilities, the probability that in an encounter between two individuals of different classes 
the one or the other is paying, the tax rates relative to the different income classes and the percentages of evasion. 

\smallskip

They
take the form
\begin{equation}
{{d x_j^{\alpha}} \over {d t}} =  
\sum_{h,k=1}^{n} \sum_{\beta,\gamma=1}^{m} {\Big (} C_{{(h,\beta)};{(k,\gamma)}}^{(j,\alpha)} 
+ T_{[{(h,\beta)};{(k,\gamma)}]}^{(j,\alpha)}(x) {\Big )}
x_h^{\beta} x_k^{\gamma}     -    x_j^{\alpha}  \sum_{k=1}^{n}  \sum_{\gamma=1}^{m} x_k^{\gamma} \, , 
\label{diffequations}
\end{equation}
for $j = 1,...,n$ and $\alpha = 1,..,m$,
where

\bigskip

$\bullet$ for any $h, k, j = 1,...,n$ and any $\alpha, \beta, \gamma = 1,...,m$, the coefficient 
$$
C_{{(h,\beta)};{(k,\gamma)}}^{(j,\alpha)} \in [0,+\infty)
$$ 
expresses the probability that an $(h,\beta)$-individual will belong to 
the group $(j,\alpha)$ as a consequence of a direct interaction with an $(k,\gamma)$-individual.

\noindent These coefficients satisfy $\sum_{j=1}^{n} \sum_{\alpha}^{m} C_{{(h,\beta)};{(k,\gamma)}}^{(j,\alpha)} = 1$ 
for any fixed $(h,\beta)$, $(k,\gamma)$;

$\bullet$ for any $h, k, j = 1,...,n$ and any $\alpha, \beta, \gamma = 1,...,m$, the function
$$
T_{[{(h,\beta)};{(k,\gamma)}]}^{(j,\alpha)}(x) : {\bf R}^{n\times m} \to {\bf R}
$$ 
expresses the variation in the group $(j,\alpha)$ 
due to the taxation and redistribution process associated to an interaction between an $(h,\beta)$-individual 
with an $(k,\gamma)$-individual. 

\noindent These functions are continuous and 
satisfy $\sum_{j=1}^{n} \sum_{\alpha}^{m}  T_{[{(h,\beta)};{(k,\gamma)}]}^{(j,\alpha)}(x) = 0$ 
for any fixed $(h,\beta)$, $(k,\gamma)$ and $x \in {\bf R}^{n\times m}$.

\smallskip

A particular choice of the $C_{{(h,\beta)};{(k,\gamma)}}^{(j,\alpha)}$'s and the $T_{[{(h,\beta)};{(k,\gamma)}]}^{(j,\alpha)}(x)$'s 
will be proposed below.
Before doing it, we need to introduce and motivate some further terms.

Similarly as in \cite{RefBM3,RefBM4} we first define for $h, k = 1, ... , n$ certain coefficients $p_{h,k}$,
aimed to specify
the probability that  in an encounter between an individual of the $h$-th  income class and one of the $k$-th class,
the one who pays is the former. 
We take
$$
p_{h,k} = \min \{r_h,r_k\}/{4 r_n} \, ,
$$
with the exception of the terms
$p_{j,j} = {r_j}/{2 r_n}$ for $j = 2, ..., n-1$,
$p_{h,1} = {r_1}/{2 r_n}$ for $h = 2, ..., n$, 
$p_{n,k} = {r_k}/{2 r_n}$ for $k = 1, ..., n-1$,
$p_{1,k} = 0$ for $k = 1, ..., n$
and
$p_{h,n} = 0$ for $h = 1, ..., n$.
The introduction of these coefficients implies that,
with only reference here to the income class (and independently of the evasion sector), for each $h - k$ pair there is 
\begin{itemize}
\item a probability denoted by $p_{h,k} \in [0,1]$ that the $h$-individual will transfer some money to the $k$-th one,
\item a probability $p_{k,h} \in [0,1]$ that the $k$-individual will transfer some money to the $h$-th one,
\item a probability $1 - p_{h,k} - p_{k,h} \in [0,1] $ that the two do not exchange money.
\end{itemize}
Correspondingly, we are assuming that for any $h\in \{1,...,n\}$ and any $k\in \{1,...,n\}$ the frequency of payment of individuals of the $h$-th income class 
to individuals of the $k$-th income class
is a fixed one. We could call this a compartmental representative agent behavior (see in this connection \cite{RefBis,RefTraGal}).

Then, we introduce $S$ (with $S << (r_{i+1} - r_{i})$ for all $i = 1, ..., n$),
the amount of money that in each direct transaction one individual is supposed to pay to another.
The individual who receives the money is expected to pay a part of this as a tax to the government.
If this individual belongs to the $k$-th income class,
we may for sake of simplicity assume that he should pay  an amount corresponding to $S \, \tau_k$,
$\tau_k$ being the tax rate of his income class.

At this point we notice that
in a tax compliance case the effect of a direct interaction with
an individual of the $h$-th income class paying $S$ to one of the $k$-th class and this paying the due tax
would be equal to that of the first individual paying an amount $S\, (1 - \tau_k)$ to the $k$-th income class one
and paying as well a quantity $S \, \tau_k$ to the government or equivalently, due to the redistribution,
to the community of individuals.\footnote{\ Actually, in this model all individuals, but those of the $n$-th income class 
benefit from the redistribution.
Differently, also individuals of the $n$-th class could advance to a higher class, but this is no possible.} 

Being interested in a tax evasion case study, we now introduce, beside the tax rates $\tau_k$, some other parameters.
For any $\alpha = 1,...,m$ let 
$$
\theta_{ev}(\alpha) \in [0,1] 
$$
denote the percentage of the due taxes payed by individuals characterized by an evasion behavior index $\alpha$.
Then, we define for any $k = 1, ..., n$ and $\alpha = 1, ..., m$,
\begin{equation}
\theta_{k,\alpha} = \theta_{ev}(\alpha) \, \tau_k \, .
\label{taueteta}
\end{equation} 
The quantity $\theta_{k,\alpha}$ 
in (\ref{taueteta}) expresses the fraction a $(k,\alpha)$-individual
actually pays as a tax; this percentage depends both on the income class represented by the index $k$
and on the evasion index $\alpha$.

\begin{example}
As an example to illustrate the situation, take $m=3$ and consider three evasion behaviors, described by
\begin{equation}
\theta_{ev}(1) = 1 \, , \qquad \theta_{ev}(2) = 1/2 \, , \qquad \theta_{ev}(3) = 1/4 \, .
\label{example of teta evasion}
\end{equation}
In such a case one would have 
\begin{itemize}
\item individuals in the first sector paying all they should and not evading at all,
\item individuals in the second sector paying half of what they should,
\item individuals in the third sector paying one quarter of what they should.
\end{itemize}
\end{example}

The coefficients $C_{{(h,\beta)};{(k,\gamma)}}^{(j,\alpha)}$'s can now be defined: the only nonzero ones among them are:
\begin{eqnarray}
\label{definition of Cj+1}
C_{{(j+1,\alpha)};{(k,\beta)}}^{(j,\alpha)} & = 
                  & p_{j+1,k} \, \frac{S \, (1-\theta_{k,\beta}) }{r_{j+1} - r_{j}} \, , \\
\label{definition of Cj}
C_{{(j,\alpha)};{(k,\beta)}}^{(j,\alpha)} & = 
            & 1 
               - \, p_{k,j} \, \frac{S \, (1-\theta_{j,\alpha})}{r_{j+1} - r_{j}} 
               - \, p_{j,k} \, \frac{S \, (1-\theta_{k,\beta})}{r_{j} - r_{j-1}} \, ,  \\
\label{definition of Cj-1}              
C_{{(j-1,\alpha)};{(k,\beta)}}^{(j,\alpha)} & = 
               & p_{k,j-1} \, \frac{S \, (1-\theta_{j-1,\alpha})}{r_{j} - r_{j-1}} \, . 
\end{eqnarray} 
We point out that 

- in $(\ref{definition of Cj+1})$, $C_{{(j+1,\alpha)};{(k,\beta)}}^{(j,\alpha)}$ is defined only for $j \le n-1$ and $k\le n-1$;

- in the expression of $C_{{(j,\alpha)};{(k,\beta)}}^{(j,\alpha)}$ in 
$(\ref{definition of Cj})$, the second addendum is present only 
for $j \le n-1$ and $k \ge 2$, whereas the third addendum is present
only for $j \ge 2$ and $k \le n-1$; 

- in $(\ref{definition of Cj-1})$, $C_{{(j-1,\alpha)};{(k,\beta)}}^{(j,\alpha)}$ is defined only for $j \ge 2$ and $k\ge 2$;

- in $(\ref{definition of Cj+1})-(\ref{definition of Cj-1})$, the indices $\alpha$ and $\beta$ take any value in $\{1, ..., m\}$.

\smallskip

We also emphasize that the coefficients $p_{h,k}$ enter in the formulae 
$(\ref{definition of Cj+1}), (\ref{definition of Cj}), (\ref{definition of Cj-1})$ 
in such a way that their effect can be also interpreted as 
weighting the amount of money
exchanged. In other words, the situation is the same one would have assuming the frequency of payment of individuals independent on the income class,
but with the amount of money paid in each transaction by
individuals of the $h$-th income class 
to individuals of the $k$-th income class
equal to $p_{h,k} S$ instead of $S$.
The specific choice of $p_{h,k}$ adopted here
is suggested by the phenomenological observation that typically
poor people pay and earn less than rich people.

\smallskip

We take the functions $T_{[{(h,\beta)};{(k,\gamma)}]}^{(j,\alpha)}(x)$ as 
\begin{eqnarray}
\label{definition of T}
T_{[{(h,\beta)};{(k,\gamma)}]}^{(j,\alpha)}(x) 
& = & 
\frac{p_{h,k} \, S \, \theta_{k,\gamma}} {\sum_{i=1}^{n} \sum_{\lambda=1}^{m} x_{i}^{\lambda}} {\bigg (}  \frac{x_{j-1}^{\alpha}}{(r_j - r_{j-1})} 
-   
\frac{x_{j}^{\alpha}}{(r_{j+1} - r_{j})} {\bigg )} \\
\ & + &  p_{h,k} \, S \, \theta_{k,\gamma} \, 
{\bigg (} 
\frac{\delta_{h,j+1}{\delta_{\alpha,\beta}}}{r_h - r_{j}} \, - \, \frac{\delta_{h,j}{\delta_{\alpha,\beta}}}{r_h - r_{j-1}}
{\bigg )} 
\, \frac{{\sum_{i=1}^{n-1} \sum_{\lambda=1}^{m} x_{i}^{\lambda}}}{{\sum_{i=1}^{n} \sum_{\lambda=1}^{m} x_{i}^{\lambda}}} \, ,  \nonumber 
\end{eqnarray} 
with $\delta_{i,j}$ denoting the {\it Kronecker delta}. In the r.h.s. of $(\ref{definition of T})$, $h >1$ and the terms 
involving the index $j-1$ [respectively, $j+1$] are effectively present only for $j-1 \ge 1$ [respectively, $j+1 \le n$].  
The indices $\alpha$, $\beta$ and $\gamma$ take any value in $\{1, ..., m\}$.

\bigskip

{\noindent {\it {Remark 1}}}
It may be helpful stressing here the fact that, 
even if 
the equations $(1)$ describe migrations of aggregate fractions of individuals,
in fact 
a probabilistic micro-interaction modelling 
underlies the dynamical process expressed by these equations.

The right hand sides of $(1)$ contain quadratic [and other nonlinear] terms,
exactly because they give account of a large number of 
pairwise interactions [and, through the taxation and redistribution process, also of interactions involving more individuals].
For example, the origin of  the coefficients $C_{{(h,\beta)};{(k,\gamma)}}^{(j,\alpha)}$ (which refer to direct money exchange) is the following.
The interaction between 
an $(h,\alpha)$-individual and a $(k,\beta)$-individual with the 
$(h,\alpha)$-individual paying, produces the variation of the fraction of individuals in some groups (in general, four).
Indeed, the $(h,\alpha)$-individual becomes a little bit poorer, inducing a partial migration from the $h$-th income class to the $(h-1)$-th one
and the $(k,\beta)$-individual) becomes a little bit richer, inducing a partial migration from the $k$-th income class to the $(k+1)$-th one.
The mentioned variation in the four groups is described through the coefficients: 
\begin{eqnarray}
b_{{(h,\alpha)};{(k,\beta)}}^{(h-1,\alpha)} & = & p_{h,k} \, S \, (1-\theta_{k,\beta}) \, \frac{1}{r_h - r_{h-1}} \, , \nonumber \\
b_{{(h,\alpha)};{(k,\beta)}}^{(h,\alpha)} & = & - p_{h,k} \, S \, (1-\theta_{k,\beta}) \, \frac{1}{r_h - r_{h-1}} \, , \nonumber \\
b_{{(k,\beta)};{(h,\alpha)}}^{(k+1,\beta)} & = & p_{h,k} \, S \, (1-\theta_{k,\beta}) \, \frac{1}{r_{k+1} - r_{k}}\, , \nonumber \\
b_{{(k,\beta)};{(h,\alpha)}}^{(k,\beta)} & = & - p_{h,k} \, S \, (1-\theta_{k,\beta}) \, \frac{1}{r_{k+1} - 
r_{k}} \, . 
\label{bhk}
\end{eqnarray} 
The coefficients in the formulae $(\ref{definition of Cj+1})-(\ref{definition of Cj-1})$, namely those appearing in the equation $(1)$
and which refer to the $(j,\alpha)$ group, 
are then obtained from these, observing that each probability $C_{{(h,\beta)};{(k,\gamma)}}^{(j,\alpha)}$ can be written as a sum 
$$
C_{{(h,\beta)};{(k,\gamma)}}^{(j,\alpha)} = a_{{(h,\beta)};{(k,\gamma)}}^{(j,\alpha)} + b_{{(h,\beta)};{(k,\gamma)}}^{(j,\alpha)} \, ,
$$
where the 
``absence-of-variation term"
$a_{{(h,\beta)};{(k,\gamma)}}^{(j,\alpha)} = 1$ only if $j=h$ and $\alpha=\beta$, independently of $(k,\gamma)$,
and the
$b_{{(h,\beta)};{(k,\gamma)}}^{(j,\alpha)}$ are as in $(\ref{bhk})$ (see \cite{RefB} for a similar discussion in a simpler case).

We also observe that the structure of the 
$C_{{(h,\beta)};{(k,\gamma)}}^{(j,\alpha)}$ and the $T_{[{(h,\beta)};{(k,\gamma)}]}^{(j,\alpha)}(x)$
in $(\ref{definition of Cj+1})-(\ref{definition of Cj-1})$ and $(\ref{definition of T})$
is
determined by the conservation requirements of the global mechanism. The 
stochastic character they enjoy is
due to the presence of the
coefficients $p_{h,k}$ which can be defined with some degrees of freedom.
It is in view of the coefficients $p_{h,k}$ that we call this a
{\it{probabilistic micro-interaction}} modelling.
We do not know who exactly is going to interact with whom:
we only know at a probabilistic level how often
individuals in a group interact with individuals of another group. 

Of course, assuming that each individual of the $h$-th income class
has the same probability $p_{h,k}$ of paying in an encounter with an individual of the $k$-th class
corresponds to attribute the same behavior 
(intended as attitude to pay)
to all pairs of individuals of the same two specific classes.
This reminds a
mean-field approach, see e.g. \cite{RefAoki,RefAokiYos}. But, in fact, the
underlying modelling just described provides 
our approach with a different characterization.\footnote{
In the Boltzmann approach to statistical mechanics, to which our approach is inspired, the variables are described by means of a probability distribution function.
The Maxwell-Boltzmann statistics gives the expected number of particles in a given volume of the phase space.
Space and velocity are continuous variables.
The discretized Boltzmann approach is somewhat more manageable, because it groups together 
particles with the same velocity or, in the socio-economic version,
individuals in the same income class.
Integrals are then replaced by sums and the number of admissible velocities or classes
can be increased as needed to ensure the precision required in any specific case. 
In comparison to the discretized Boltzmann, the mean-field approach entails a loss of information, since the evolution equation 
for each individual implies that it interacts with an average value of all the others, and the whole evolution process is self-consistent. 
In the discretized Boltzmann approach and in our model the interactions are microscopic and occur between all individuals.
Finally, if one compares a Boltzmann approach with an agent-based simulation, it is fair to say that the Boltzmann ÒagentsÓ 
have properties typical of deterministic particles, while the behavior of simulated agents can be much more flexible. 
However, the Boltzmann approach offers the advantage of a closed mathematical formulation, independent from the software.}
$\diamond$

\bigskip

By suitably 
adapting proofs, which are relative to a previous, less general, version of the model and can be found
in \cite{RefB}, 
one may check that the following properties,
amounting to
conservation in time of both the number of individuals and the global income as well, hold true.

\medskip

\noindent {\it {Property 1}}
For any initial condition $x_0 = \{{x_{0_j}^{\alpha}}\}_{j=1,...n;\alpha=1,...m}$, 
for which $x_{0_j}^{\alpha} \ge 0$ for any $j = 1, ... , n$ and $\alpha = 1, ... , m$, and 
${\sum_{j=1}^{n} \sum_{\alpha=1}^{m} x_{0_j}^{\alpha} = 1}$,
a unique solution $x(t) =  \{{x_{0_j}^{\alpha}}(t)\}_{j=1,...n;\alpha=1,...m}$ of $(\ref{diffequations})$ exists,
which is defined for all $t \in [0,+\infty)$, satisfies $x(0) = x_0$ and also
\begin{eqnarray}
& & x_{j}^{\alpha}(t) \ge 0 \ \hbox{for} \ j = 1, ... , n \ \hbox{and} \ \alpha = 1, ... , m \nonumber \\
\label{solution in the future}
& & \qquad \qquad  \hbox{and} \\
& & {\sum_{j=1}^{n} \sum_{\alpha=1}^{m} x_{j}^{\alpha}(t) = 1} \ \hbox{for all} \ t \ge 0 \, . \nonumber
\end{eqnarray}

As a consequence of this property, the expressions of the $T_{[{(h,\beta)};{(k,\gamma)}]}^{(j,\alpha)}(x) $'s 
in (\ref{definition of T}) somehow simplify
and the right hand sides of system $(\ref{diffequations})$ turn out to be in fact polynomials of degree $3$.

\smallskip

\noindent {\it {Property 2}} The scalar function
$\mu(x)=\sum_{j=1}^n r_j \sum_{\alpha=1}^{m} x_j^{\alpha}$
remains constant along each solution of system $(\ref{diffequations})$.

\medskip

In addition, the running of several numerical solutions provides evidence of the following fact.

\medskip

\noindent {\it {Property 3}} If the parameters of the model ($r_1, ..., r_n$, $S$, the $\tau_k$'s, the $\theta_{ev}(\alpha)$'s) 
and also the fraction of individuals 
with different behavior\footnote{\ The fraction of individuals with a specific evasion behavior is assumed here to be the same in each income class.} 
are fixed, if 
$\mu \in [r_1,r_n]$ is fixed, then the solutions $x(t) =  \{{x_{0_j}^{\alpha}}(t)\}_{j=1,...n;\alpha=1,...m}$ evolving from initial conditions 
$x_0 = \{{x_{0_j}^{\alpha}}\}_{j=1,...n;\alpha=1,...m}$, 
for which $x_{0_j}^{\alpha} \ge 0$ for any $j = 1, ... , n$ and $\alpha = 1, ... , m$, 
and
which satisfy
$$
{\sum_{j=1}^{n} \sum_{\alpha=1}^{m} x_{0_j}^{\alpha} = 1}
\qquad \hbox{and} \qquad
\sum_{j=1}^n r_j \sum_{\alpha=1}^{m} x_{0_j}^{\alpha} = \mu
$$
tend asymptotically to a same stationary distribution
as $t \to +\infty$.

\section{The evidence from numerical solutions}
\label{numsim}

Our specific interest here is to analyze the income evolution  in the long time limit of groups of individuals characterized by different behaviors. 
In other words, whereas our previous investigation in \cite{RefBM3,RefBM4} was especially addressed to detect the effects of 
(a different kind of) tax evasion 
on the population as a whole, here we also focus on the 
evasion effects on the different mentioned groups. 

Finding analytical solutions of
the nonlinear differential equations $(\ref{diffequations})$ is of course hopeless.
However, numerical solutions provide sufficient information.
In order to obtain them, one has to first fix several parameters. 
We take here $n=9$, $m=3$, $S=1$, $r_j = 10 \, j$ for $j = 1, ..., n$ and
assume tax rates increasing from a minimal one, $\tau_{min}$, to a maximal one, $\tau_{max}$,
according to a progressive taxation system, as given by
\begin{equation}
\tau_j = \tau_{min} +  \frac{j - 1}{n-1} \, (\tau_{max} - \tau_{min}) \, , \qquad \hbox{for} \ j = 1, ... , n \, .
\label{progressivetaxrates}
\end{equation}
Still, the values of $\tau_{min}$, $\tau_{max}$ 
and $\theta_{ev}(\alpha)$ for $\alpha = 1, ... , m$
remain to be chosen.

\medskip

\begin{figure*}
\begin{center}
\includegraphics[width=3.3cm,height=1.75cm] {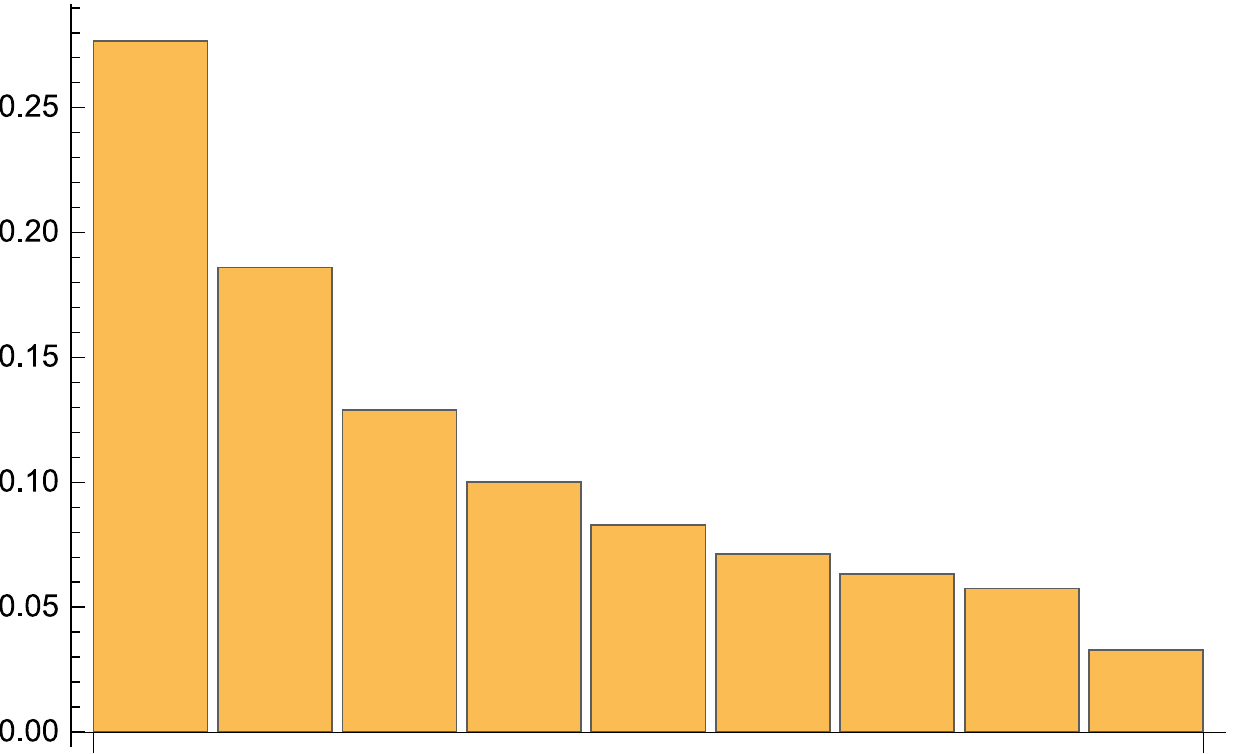}
 \hskip0.5cm
\includegraphics[width=3.3cm,height=1.75cm] {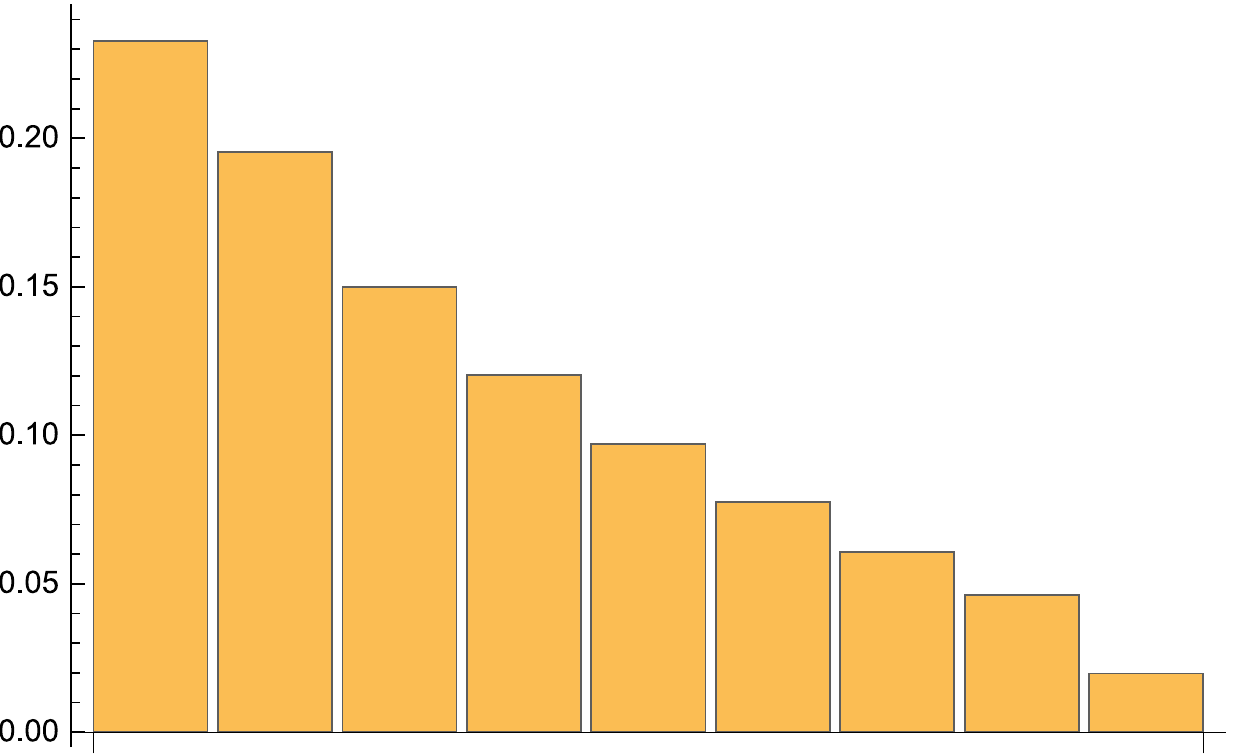}
 \hskip0.5cm
\includegraphics[width=3.3cm,height=1.75cm] {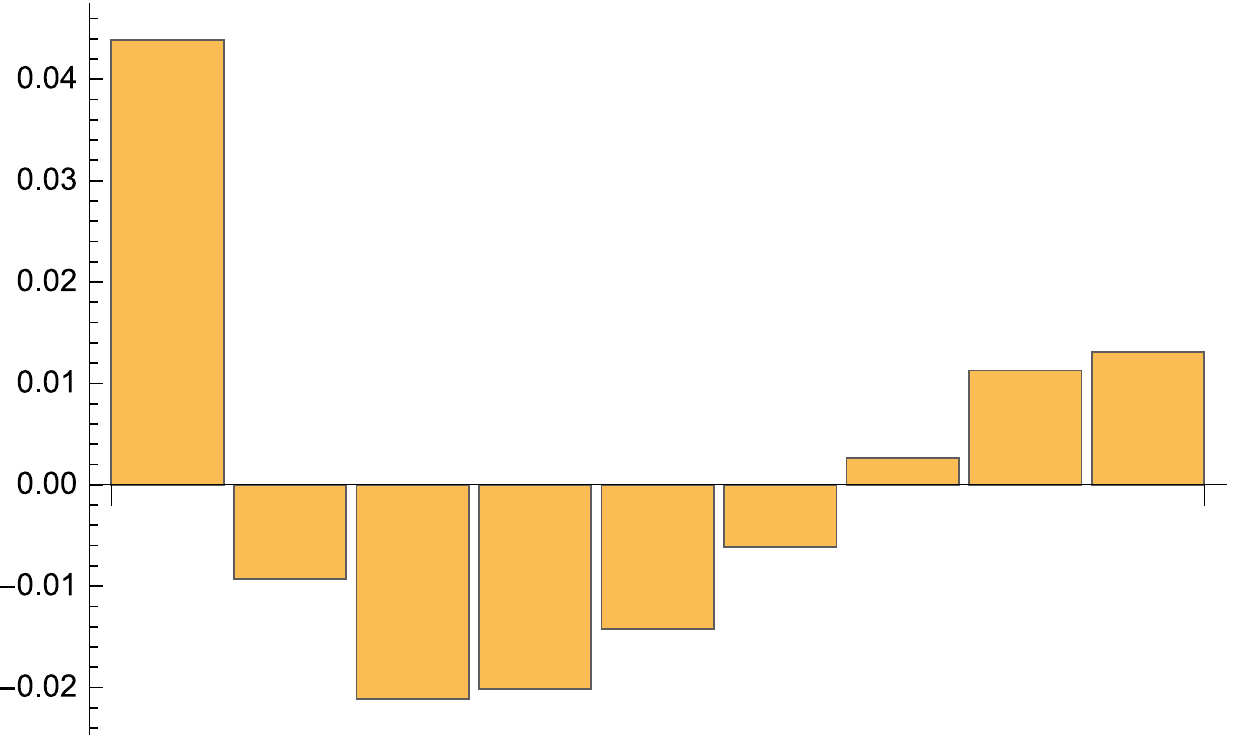}
\end{center}
\caption{Collective effect of tax evasion. The asymptotic income distribution on the left refers to a case with tax evasion, 
the one in the middle refers to a tax compliance case with the same initial conditions 
and the panel on the right displays the difference of the fraction of individuals in the various classes in the first and second case. 
Note that the figures are scaled differently.}
\label{fig:1}       
\end{figure*}

Each time we explore  aggregate effects in the asymptotic stationary income distribution 
of the population and compare situations of tax compliance and tax evasion, 
we find that evasion leads to an increment in the number of individuals in the poorest and in the richest classes, 
at the detriment of the middle classes.  Fig.\ $\ref{fig:1}$ illustrates the typical situation.
In \cite{RefBM3,RefBM4} we have shown that tax evasion has in general the effect to increase economic inequality, 
as measured for instance by the Gini index. We have also studied situations in which evasion grows 
in response to an increase of the tax rates, finding an optimal ``compromise" characterized by minimum inequality. 
These results were obtained in conditions of homogeneous evasion rates.

\smallskip

\begin{figure*}
\begin{center}
\includegraphics[width=3.3cm,height=1.75cm] {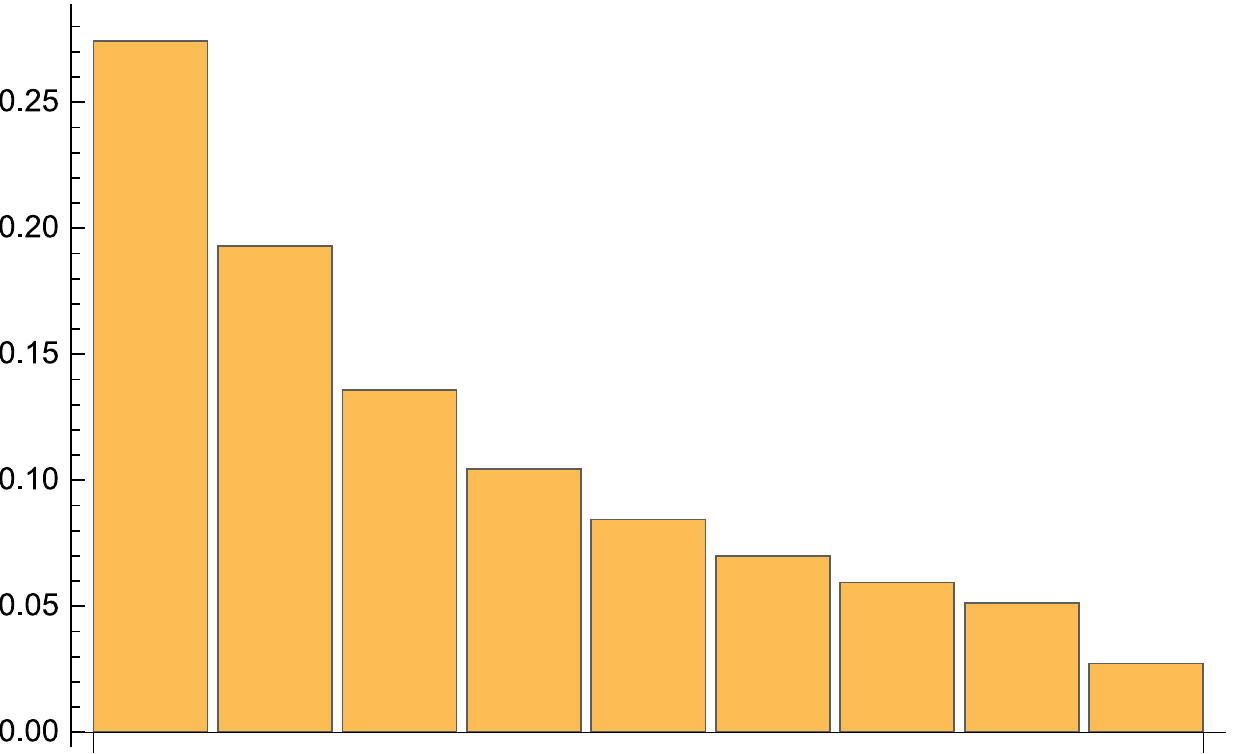}
 \hskip0.5cm
\includegraphics[width=3.3cm,height=1.75cm] {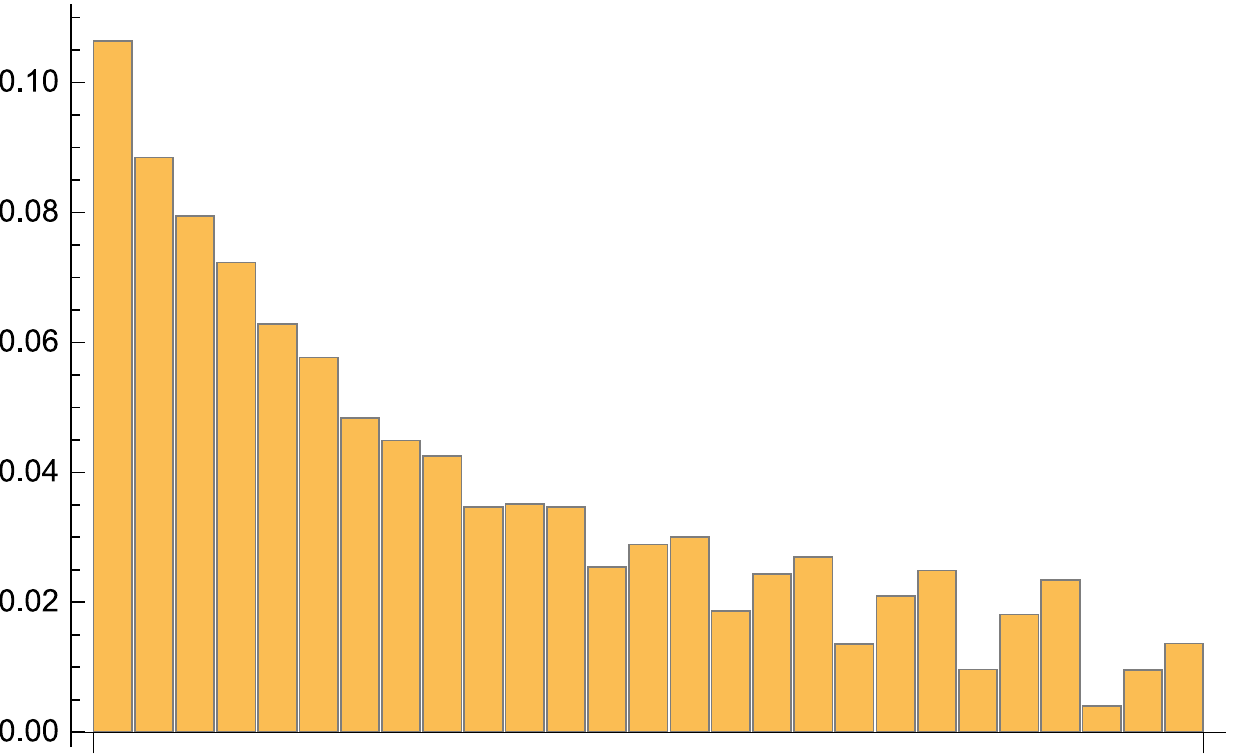}
 \hskip0.5cm
\includegraphics[width=3.3cm,height=1.75cm] {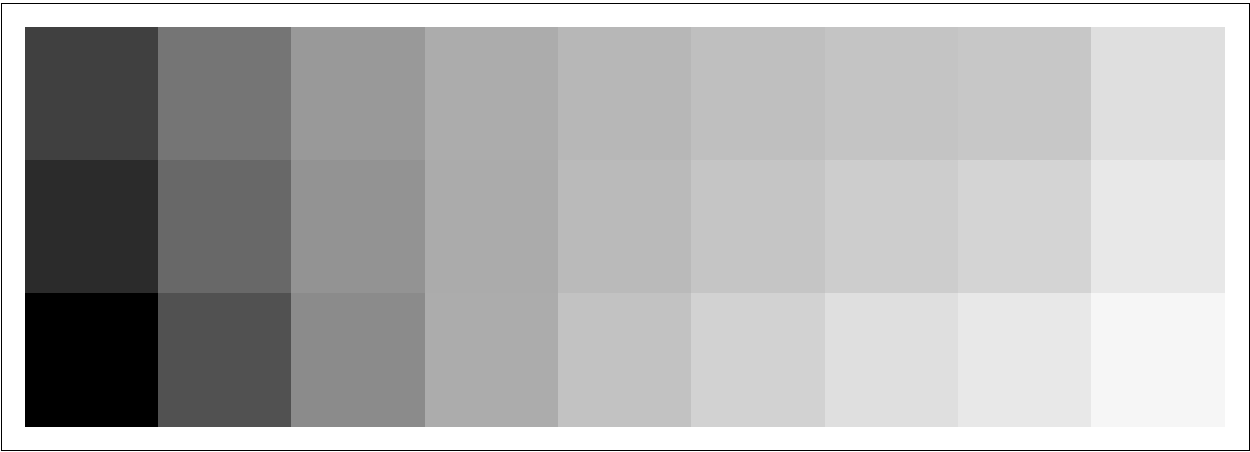}
\end{center}
\begin{center}
\includegraphics[width=3.3cm,height=1.75cm] {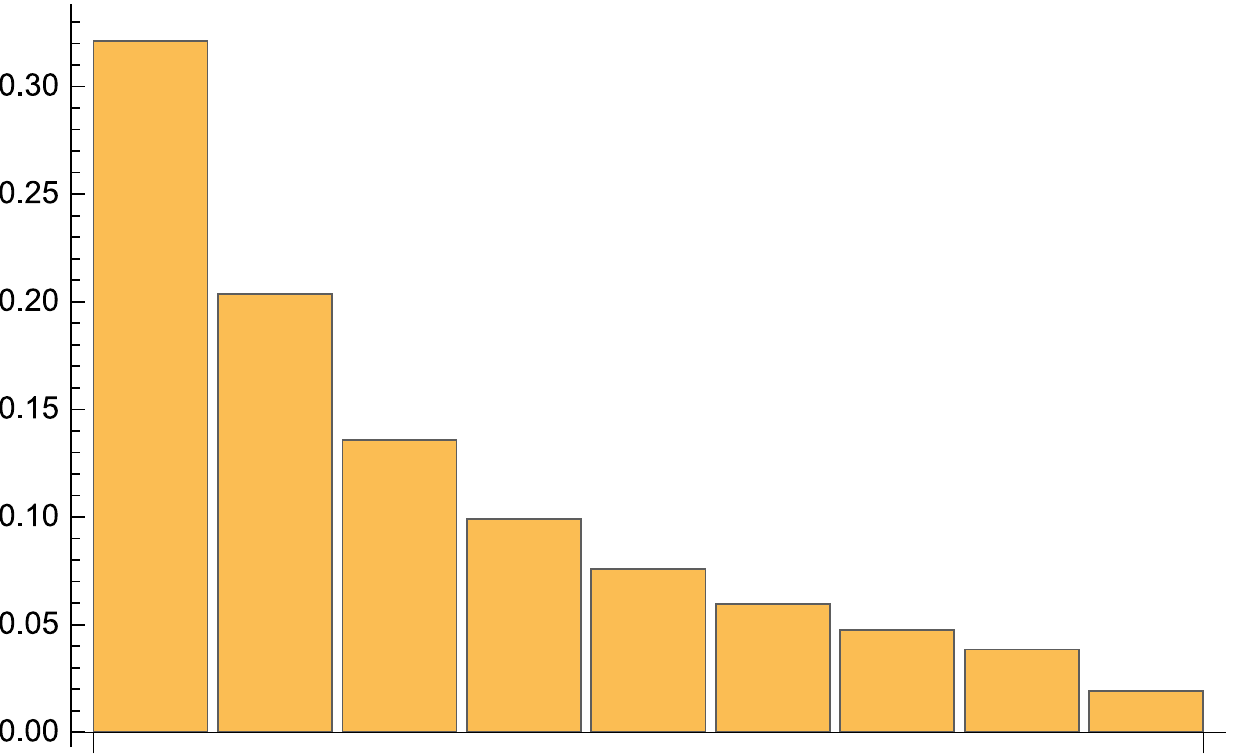}
 \hskip0.5cm
\includegraphics[width=3.3cm,height=1.75cm] {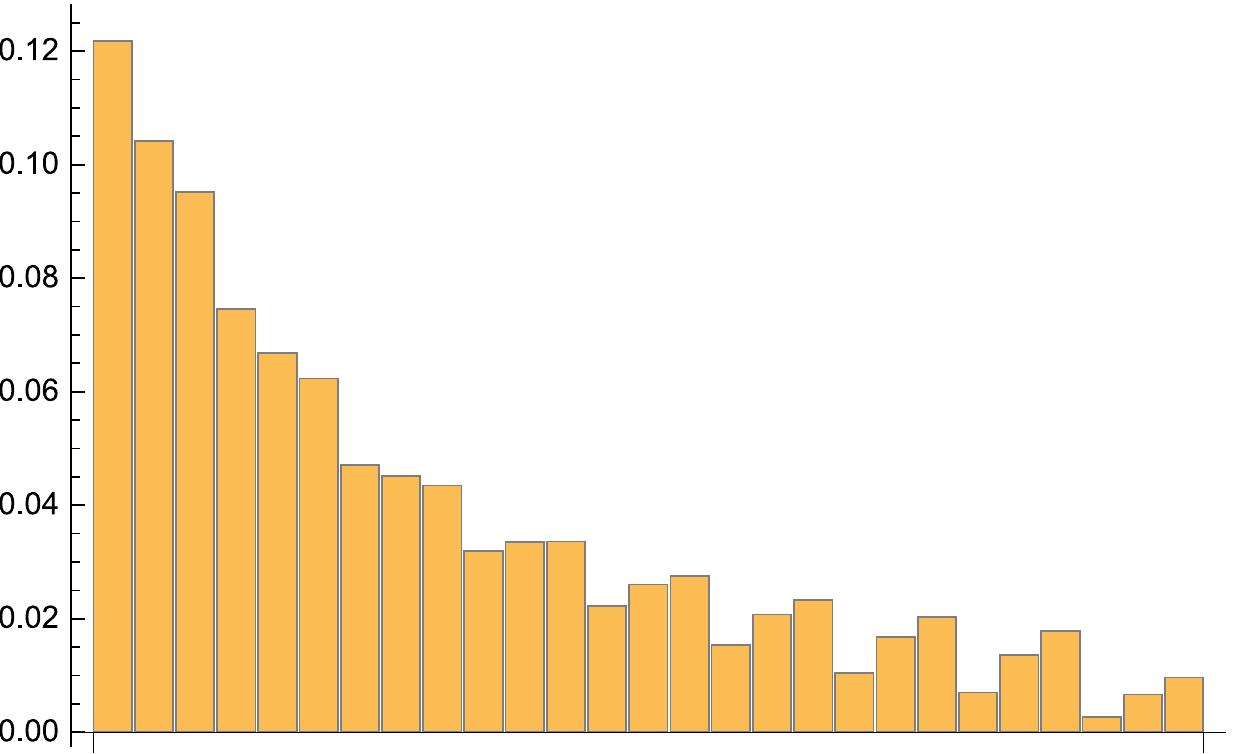}
 \hskip0.5cm
\includegraphics[width=3.3cm,height=1.75cm] {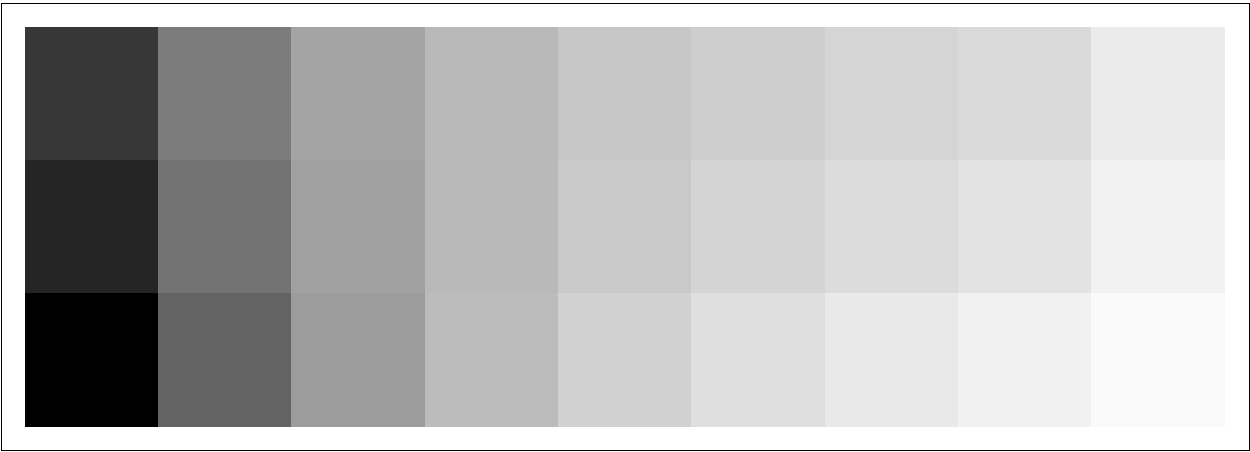}
\end{center}
\caption{In the two rows the asymptotic income distribution referring to two different initial data are plotted. 
The two panels on the left display the distributions by income classes. The histograms in the central panels represent the sequence 
for each income class of  three evasion behavior sectors with 
$\theta_{ev}(1) = 1, 1/2, 1/4$.
The panels on the right provide an alternative bi-dimensional representation, in which the sectors can also be distinguished.}
\label{fig:2}       
\end{figure*}

Assume now, to fix ideas, that the evasion behaviors are as in 
$(\ref{example of teta evasion})$ in Example $1$ and that each of them is present in one third of the population.
\footnote{\ The choice of this particular subdivision is motivated by the desire to
derive a balanced comparison of the evolution of the different groups.}
Accordingly, in each income class one third of the individuals pays all due taxes, 
one third pays half of them and one third pays one quarter of them.
Not only can we obtain information on the aggregate shape of the asymptotic income distribution.
We also get a more detailed picture of the effects of tax evasion within different behavioural sectors. 
Specifically, it turns out that in low income classes the numbers of ``honest individuals'', ``half-evaders'' and ``three-quarters evaders'', 
counted in this order, are 
decreasing from the largest to the smallest,
while the situation is reversed in the high income classes. 
A graphic illustration of this is provided by the Fig.\ $\ref{fig:2}$.

\smallskip

In order to understand which variations of the model parameters can best represent some real situations, 
and what results we can expect, we can compare
situations in which evasion is widespread with situations in which it is confined only to a part of the population, 
the total evasion level being the same. 
By total evasion level we denote the total fraction of tax payments which is eluded, 
summed over all sectors of the population. 
With three sectors, for instance, we can consider the two following situations, which have both total evasion level $1/6$, 
but exhibit respectively widespread and concentrated evasion: 

\begin{enumerate}
\item 
one sector is honest, with $\theta=\theta_{ev}=1$, i.e.\ 100\% of taxes are paid, 
while the other two sectors are slightly dishonest, with $\theta=0.75$, i.e.\ 75\% of taxes are paid; 
\item 
two sectors are honest and the third is quite dishonest, with $\theta=0.5$. 
\end{enumerate}

We wonder whether the Gini index is significantly different in the two cases, or in other words, 
if evasion has a different impact on inequality when it is widespread or confined to a part of the population. 
Furthermore, we might wonder whether the partial Gini indices of the behavioral sectors are substantially different 
from that of the total population, for instance in the sense that among evaders there is more inequality than among honest taxpayers. 
The numerical solutions show, however, that in all these cases the Gini index does not exhibit any significant variations, 
depending only on the total evasion level.
 
\smallskip
 
The introduction of behavioral sectors into the model allows to observe another important effect, 
namely the appearance of a clear difference between the average incomes of honest taxpayers and tax evaders. 
It is already apparent from the income histograms in Fig.\ $\ref{fig:2}$
that, as expected, tax evaders tend to get richer than honest taxpayers:
the bars representing the lowest-income classes clearly show a larger share of honest people, 
while the opposite happens for the highest-income classes, where more evaders are found. 
It is interesting to give a quantitative estimate of this difference and to study its dependence on the evasion level. 

To this end, let us consider the relative difference $d$ between the average income 
of the worst evaders and the average income of the honest taxpayers: $d=(\mu_m-\mu_1)/\mu_1$. 
This relative difference is quite large, typically of the order of 10 to 20\% in the histograms of Fig.s \ref{fig:1}, \ref{fig:2}. 
In order to evaluate its dependence on the evasion level, we need to choose a fixed and reasonable ``evasion spread" pattern. 

\begin{table}
\caption{Percentage gap $d$ between the average income 
of the worst evaders and the average income of honest taxpayers in dependence of the total evasion level $\eta$. 
A gradual spread of evasion in three behavioral sectors is assumed. 
For instance, when the total evasion level is 10\% (the first sector pays 100\% of the due taxes, the second pays 90\% and the third 80\%), 
the average income of the third sector is 6.8\% larger than the income of the first sector. 
The applied tax rates grow linearly with income between $\tau_{min}=10\%$ and $\tau_{max}=45\%$. 
The dependence $d(\eta)$ can be approx.\ fitted as $d(\eta) \simeq 0.42 \eta^2+0.62 \eta$.}
\label{tab:1}
\begin{tabular}{ccc}
\hline\noalign{\smallskip}
Tot.\ \% evasion level $\eta$ & \% of due taxes paid in the three sectors & \% income gap $d$  \\
\noalign{\smallskip}\hline\noalign{\smallskip}
    5 & 100, \ 95, \ 90 &  3.5 \\
  10 & 100, \ 90, \ 80 &  6.8 \\
  15 & 100, \ 85, \ 70 &  10.8 \\
  20 & 100, \ 80, \ 60 &  14.6 \\
  25 & 100, \ 75, \ 50 &  18.1 \\
  30 & 100, \ 70, \ 40 &  21.5 \\
  40 & 100, \ 60, \ 20 &  31.8 \\
  50 & 100, \ 50, \ \ 0 &  41.8 \\

\noalign{\smallskip}\hline
\end{tabular}
\end{table}

We can proceed as follows: 
suppose that the first sector is always honest ($\theta=1$), the second sector has $\theta=1-\eta$ (evasion level $\eta$), 
and the third sector has $\theta=1-2\eta$ (evasion level $2\eta$). The total evasion level is $3\eta/3=\eta$. 
Let us gradually increase $\eta$, from 0 to 0.5, and compute $d(\eta)$. Results are shown in Tab.\ \ref{tab:1}.

The dependence is manifestly non-linear, showing that the phenomenon is complex and its interpretation not simple. 
In fact, an increase in the evasion level affects the income distribution in at least two ways: (a) through direct interactions, 
because evaders gain systematically more from any interaction; (b) through indirect interactions (tax redistribution), 
because when tax evaders tend to outnumber honest taxpayers in the higher-income classes, which should pay higher tax rates, 
the total tax collection is diminished. It is not obvious, however, that this diminution should further increase the difference $d(\eta)$ 
defined above, since in the present version of the model redistribution is uniform. In the version with heterogeneous redistribution \cite{RefBM5}
we could take into account more subtle real effects: for instance, supposing that welfare provisions are means-tested, 
we could assign them based on the tax paid, instead than on the real income. 
This is known to give a further unjust and detestable advantage to tax evaders. We plan to address these issues in future work.

\section{Concluding remarks and further perspectives}
\label{conclud}

In this paper, a kinetic-type model describing economic interactions, taxation and redistribution in a closed society 
is discussed. The focus is on the effects produced by tax evasion by individuals who 
 under-declare their income in different measure.

The model suggests the following considerations.
From the sound point of view of individuals who care for society, tax compliance plays an important role towards the overcoming of economic inequality.
From the point of view of selfish individuals, the probability of improving their own economic status is higher when evading. 
The result is not surprising.
Certainly, it has to be noticed that no audit actions or punishment are taken here into consideration. 
Incorporating them in the model and investigating possible impacts would be an interesting point to explore in future work.

Another, related, point which deserves further work is the 
necessity of inclusion into the model of possible
changes of the behavioral taxpayer attitudes.
Of course, in real life, the propensity of individuals to be compliant or non compliant
does not remain always constant in time:
in particular, it
can be influenced by the behavior of others and by specific experiences (such as fines)
with fiscal agencies.
Combining into the model
the treatment of the money distributional aspect 
with behavioral and psychological factors
remains a major challenge to be faced in the future.
As well, the possibility of bankruptcy and production should be taken into account
in an effort of an improved realism.

In this connection and in conclusion, we like to emphasize that 
our goal here has been mainly a methodological one.
The interest of this model 
(possibly bound to be further enhanced)
lies in our opinion in its possible explorative use: 
simulations corresponding to different conceivable parameters allow to understand and forecast 
the emergence of different scenarios and could possibly suggest policies addressed to favour desired trends or prevent  undesired ones.

\bigskip

{\bf{Acknowledgement}}

We thank the referee for a very careful review and several constructive comments.

\end{document}